# Extreme spin-orbit coupling in Hermite-Gaussian beams in a uniaxial crystal

**T. Fadeyeva, K. Kotlyarov, A. Volyar**
*Taurida National University, 95007,Ukraine, Simferopol, Vernadsky av. 4*
*e-mail: volyar@crimea.edu*

Transformation of the spin and orbital angular momentum of Hermite-Gaussian beams of a complex and real argument propagating through a uniaxial crystal is considered. We revealed that the spin and orbital angular momentum of the complex argument HG beam experience a sharp splashes whereas such an intense conversion process is not inherent in a real argument HG beams. The reasons and conditions of the resonance effect are brought to light.

It is well known that Hermite-Gaussian beams (HG) propagating through free space or a homogeneous isotropic medium cannot carry over an orbital angular momentum (OAM) [1] while their spin angular momentum (SAM) is defined by the initial polarization states [2]. On the other hand, a traditional lens converter [3] enables us to transform HG beam into a singular one bearing optical vortices that possess the OAM. However, presence of optical vortices in the beam is not the exclusive requirement. For example, even a fundamental Gaussian beam subjected to an astigmatic transformation gains also the OAM [4]. Transmitting through a homogeneous isotropic medium, light beam changes neither the OAM nor SAM. Absolutely other situation appears in a non-homogeneous isotropic medium, for example, in an optical fiber [5, 6]. Polarization transformations and change of a vortex structure in the optical fiber are controlled by a spin-orbital coupling, a total angular momentum flux along the fiber axis being conserved. A sharp gradient of a refractive index causes an intense spin-orbit coupling resulting in essential field transformations such as, for example, a transverse shift of the reflected and refracted beams on the boundary face of two homogeneous isotropic media [7].

A circularly polarized Gaussian beam propagating along the optical axis of a homogeneous uniaxial crystal gets also the OAM owing to nucleation of a double-charged optical vortex in the orthogonally polarized beam component [8,9]. Ciattoni et.al.[10] showed that such a vortex nucleation in an anisotropic medium is a result of a spin-orbit coupling on the stipulation that a total angular momentum flux along the crystal optical axis is conserved.. A gradual beam depolarization in the Gaussian beam is accompanied by a slow growth of the OAM so that the OAM can reach a maximum value only at infinitely large crystal length. By analyzing this situation we supposed that an intense spin-orbit coupling at a relatively small crystal length can be in line with sharp intensity variations over the beam cross-section. Such characteristic features have a whole family of high-order beams: Hermite-Gaussian, Laguerre-Gaussian, Bessel-Gaussian at alias.

The aim of this Letter is to consider intrinsic features of a spin-orbital coupling in Hermite-Gaussian beams propagating along the optical axis of a homogeneous uniaxial crystal and to bring to light the conditions for the resonance transformations of the spin and orbital AM.

Let us consider the propagation of a circularly polarized HG beam along the optical axis of a uniaxial crystal with a permittivity tensor in diagonal form: $\hat{\varepsilon} = diag(\varepsilon_o, \varepsilon_o, \varepsilon_3)$, where $n_o = \sqrt{\varepsilon_o}$ and $n_3 = \sqrt{\varepsilon_3}$ being the refractive indices along a major crystallographic axes (see Fig.1). The basic equation for the transverse components of the complex amplitudes $\tilde{\mathbf{E}}_\perp = (\mathbf{e}_+ \tilde{E}_+ + \mathbf{e}_- \tilde{E}_-)$ (where $\mathbf{e}_+$ and $\mathbf{e}_-$ being the unit vectors of a circularly polarized basis) of the paraxial beam field $\mathbf{E}_\perp = \tilde{\mathbf{E}}_\perp(x,y,z)\exp(-ik_o z)$ (where $k_o = k_0 n_o$, $k_0$ stands for a wavenumber in free space) is presented in the paper [9]. Generally speaking, there are two types of HG beams [11]: the HG beams of a real argument $\mathcal{H}_{m,n}$ (or standard beams) and the HG

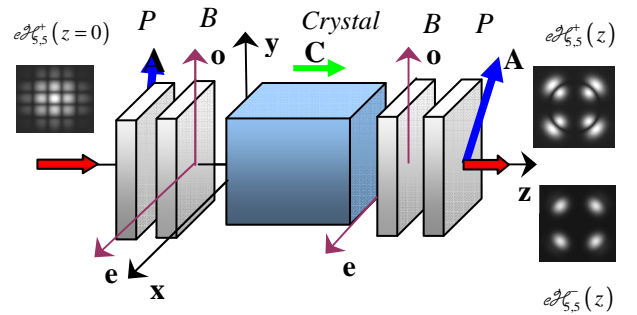

Fig.1 (on-line) Sketch of the beam propagation through the crystal and polarization filters.: *P*- a polarizer, *B* – a quarter-wave retarder, **A** – a transmittance axis of the polarizer, **C** – a unit vector of the crystal optical axis, e and o – crystallographic axes of the birefringent phase retarder and the intensity distributions of the circularly polarized components of the HG beam with m=n=4, $w_0 = 8\,\mu m$; z=4.7 mm

beams of a complex argument $e\mathcal{H}_{m,n}$ (or elegant HG beams):

$$\mathcal{H}_{m,n}^{(o,e)}(x,y,z) = H_m\left(\sqrt{2}X/|\sigma_{o,e}|\right) \times H_n\left(\sqrt{2}Y/|\sigma_{o,e}|\right)e^{i(m+n)\Gamma_{o,e}}\Psi_0^{(o,e)}, \quad (2)$$

$$e\mathcal{H}_{m,n}^{(o,e)}(x,y,z) = \left(\sqrt{\sigma_{o,e}}\right)^{-m-n} \times H_m\left(X/\sqrt{\sigma_{o,e}}\right)H_n\left(Y/\sqrt{\sigma_{o,e}}\right)\Psi_0^{(o,e)}, \quad (3)$$

where $H_m(x)$ stands for a Hermite polynomial, $\Psi_0^{(o,e)} \exp\{-R^2\sigma_{o,e}\}/\sigma_{o,e}$, $\sigma_o = 1 - iz/z_o$, $z_o = k_o w_0^2/2$, $\sigma_e = 1 - iz/z_e$, $z_e = k_e w_0^2/2$, $k_e = k_0 n_e$, $\tan\Gamma_{o,e} = z/z_{o,e}$, $n_e$ being the refractive index of the extraordinary beam, $w_0$ is a beam waist radius at the $z=0$ plane, indices o- and e- designate the ordinary or extraordinary beam in the crystal, respectively. We suppose that the electric field $\tilde{\mathbf{E}}_\perp$ of the HG beam is right hand polarized at the $z=0$ plane:

$$\tilde{\mathbf{E}}_\perp^{(m,n)}(x,y,z=0) = \mathbf{e}_+ e\mathcal{H}_{m,n}^{(o,e)}(x,y,z=0) \quad \text{(or}$$

$$\tilde{\mathbf{E}}_\perp^{(m,n)}(x,y,z=0) = \mathbf{e}_+ \mathcal{H}_{m,n}^{(o,e)}(x,y,z=0)). \quad \text{A}$$

particular solution to the paraxial wave equation (1) in terms of $e\mathcal{H}_{m,n}$ mode beams of a complex argument can be presented as

$$\tilde{E}_+ = e\mathcal{H}_+^{(m,n)} = \left[e\mathcal{H}_{m,n}^{(o)} + e\mathcal{H}_{m,n}^{(e)}\right]/\sqrt{2},$$

$$\tilde{E}_- = e\mathcal{H}_-^{(m,n)} = \frac{(-w_0)^{m+n}}{\sqrt{2}}\frac{\partial^{m+n}}{\partial x^m \partial y^m}\mathcal{G}, \quad (4)$$

where $\mathcal{G} = -e^{i2\varphi}\left[(\sigma_o\Psi_o - \sigma_e\Psi_e)/R^2 + \Psi_o - \Psi_e\right]$, $n_e = n_3^2/n_o$, $R^2 = X^2 + Y^2$, $\tan\varphi = y/x$. The field equations for the standard $\mathcal{H}$-beams can be rewritten in terms of the complex amplitude $e\mathcal{H}$-beams by means of the expression

$$\mathcal{H}_{2n+s}(x) = a\sum_{k=0}^{n} 2^k/[(n-k)!(2k+s)!]e\mathcal{H}_{2n+s}(x) \quad (5)$$

and $s = 0,1$, $a = 2^{s/2}(2n+s)!$. The major difference between the HG beams of the real and complex argument is that the standard $\mathcal{H}$-beams do not change their structure when propagating in free space up to the scale. Structure of the elegant $e\mathcal{H}$-beam is radically transformed over all length of propagation. A little displacement of the beam cross-section from the initial plane $z=0$ results in vanishing edge dislocations in the $e\mathcal{H}$-beam while the edge dislocations in the $\mathcal{H}$-beam are slightly shifted (see Fig. 2). As the $e\mathcal{H}$-beam propagates along the crystal, the beam field is concentrated around four symmetric spots (see the intensity distributions in the dotted frame in Fig.2) provided that $m=n$ regardless of the magnitude of the indices. However, the sizes of the spots decrease when enlarging the index value. At a large crystal length, the spots are modulated by a set of interference fringes. Naturally, such a beam reconstruction cannot but have an impact on the spin-orbit coupling.

In the paraxial approximation, the conservation law of the AM flux along the crystal optical axis can be presented as [10]

$$\ell_z(z) + s_z(z) = \ell_z(z=0) + s_z(z=0) = const, \quad (6)$$

where the OAM $\ell_z$ and the SAM $s_z$ are

$$\ell_z = -iK\langle\mathbf{E}|\hat{\ell}_z|\mathbf{E}^*\rangle, s_z = iK\int_\infty (E_+ E_+^* - E_- E_-^*)dS, \quad (7)$$

$K = const$, $\hat{\ell}_z = x\partial/\partial y - y\partial/\partial x$. Taking into account eqs (4), the forth integration of eqs (7) gives the expression for SAM and OAN in the $e\mathcal{H}$-beams:

$$s_z = A\cos(N\gamma), \quad \ell_z = 1 - s_z, \quad (9)$$

$$A = \left(1 + Z^2\right)^{-(m+n+1)/2} \quad N = (m+n+1), \tan\gamma = Z,$$

$Z = (1/z_o - 1/z_e)z/2$. A similar expression for the real argument $\mathcal{H}$-beams has a cumbersome form and is not presented in the Letter. The above expressions are of a generalization of the expression (52) in the paper [10] obtained for a circularly polarized fundamental Gaussian beam. Notice that the SAM and OAM fluxes vanish for the linearly polarized initial beam. A set of curves shown in Fig.3 describes evolution of the SAM $s_z$ and OAM $\ell_z$ along the crystal length. In contrast to a smooth

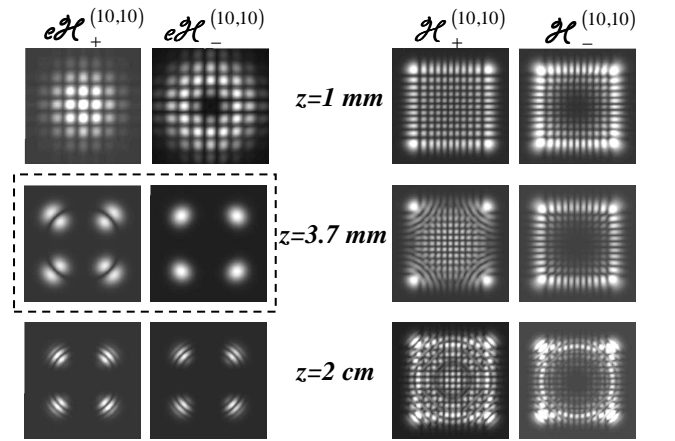

Fig. 2 Transformation of the circularly polarized components in the $e\mathcal{H}_{10,10}$ and $\mathcal{H}_{10,10}$ beams along the $LiNbO_3$ crystal length, $w_0 = 10\mu m$

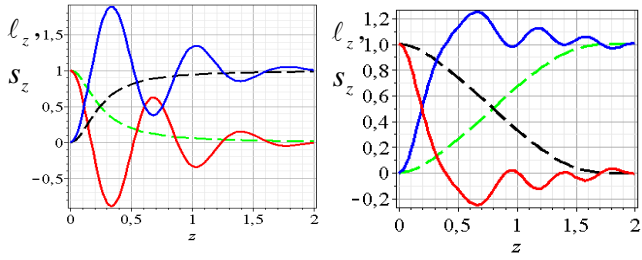
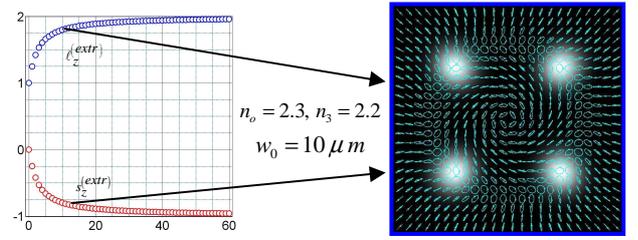

Fig.3 (on-line) Normalized SAM $s_z = S_z/K$ and OAM $l_z = L_z/K$ as functions of the length $Z$

curves for a Gaussian beam (m-n=0), the SAM and OAM of the complex argument $e\mathcal{H}$-beams experience a sharp spike near an optimal crystal length $Z = Z_{opt}$ where the AM reaches the extreme value $s_z = s_{extr}$, $l_z = l_{extr}$. A magnitude of the AM splashes depends on the beam indices m and n: the more the index value, the more the more modulus of the spin and orbital AM. However, a total AM is conserved. At the same time, the SAM and OAM for the real argument $\mathcal{H}$-beams have more smooth behavior. For example, the $\mathcal{H}$-beam with indices m=n=50 has the extreme OAM less than $l_{extr} < 1.2$ while the $e\mathcal{H}$-beam even for the indices m=n=10 gains the extreme OAM $l_{extr} \approx 1.8$.

The equation for the optimal $Z_{opt}$ can be derived from the requirement: $d s_z / d z = 0$. After a simple algebra we come to the characteristic equation: $\tan\left[(m+n+1)\arctan Z\right] = -Z$. The second root of this equation corresponds to the optimal crystal length $Z = Z_{opt}$. When enlarging the indices $m$ and $n$, the optimal length $Z_{opt}$ tends to zero: $Z_{opt} \ll 1$ and we can write approximate solution to the above equation: $(m+n+1)Z_{opt} - \pi \approx -Z_{opt}$, $m+n \gg 1$ or $Z_{opt} \approx \pi/(m+n+2)$. Fig.4,a illustrates the extreme magnitudes of the AM as a function of the indices m=n. Noteworthy that the extreme OAM reaches $l_{extr} \approx 2$ for $m > 40$ and then changes very slowly.

There are three physical processes in the crystal that underline the extreme spikes of OAM. First of all, when transmitting, high-order $e\mathcal{H}$-beams are transformed in such a way that light forms only four symmetric maxima regardless of the beam indices. The second, a uniaxial crystal forms a non-uniformly polarized field distribution in the on-axis propagating beam consisting of interlaced rings of right- and left-hand circular polarization. Typical map of the field distribution in $e\mathcal{H}$ beam illustrates Fig.4,b. The left hand circular polarization

Fig. 4 (a) Extreme values of the SAM $s_z^{(extr)}$ and OAM $l_z^{(extr)}$ as a function of the index m: m=n, and (b) The map of the polarization state distribution on the background of the total intensity distribution in the $e\mathcal{H}_{m,n}$ beam with m=n=10, Z=0.0785.

positioned at the central parts of all four intensity maxima of the beam results in arising a black strip in the $e\mathcal{H}_+$ component in Fig.2 (the framed pictures). The left hand circular polarized component dominates in the beam i.e. the SAM changes a sign. The resonance condition corresponds to coinciding the radius of the intensity maxima of the beam and the radius of the ring with a left hand circular polarization. A conservation law of the total AM flux causes a slash of the OAM at the expense of a spin-orbit coupling in a homogeneous anisotropic medium of a uniaxial crystal.

The authors thanks E. Abramochkin for a fruitful discussion of the complex and real argument HG beams.